\documentclass[12pt]{iopart}
\usepackage{graphicx}

\usepackage{anyfontsize}
\usepackage{lmodern}
\usepackage{bm}
\usepackage{amssymb}
\usepackage[colorlinks, citecolor = blue, urlcolor = blue]{hyperref}

\usepackage{mathrsfs}
\usepackage{mathtools} 
\usepackage{color}
\usepackage{verbatim}

\def\ba{\begin{eqnarray}}
\def\ea{\end{eqnarray}}
\def\be{\begin{equation}}
\def\ee{\end{equation}}

\begin{document}

\title{Lyapunov-Schmidt bifurcation analysis of a supported compressible elastic beam}

\author{
Ee Hou Yong$^{1,\ast}$ and L. Mahadevan$^{2,3,4,\dagger }$}

\address{$^1$ Division of Physics and Applied Physics, School of Physical and Mathematical Sciences, Nanyang Technological University, Singapore 637371, Singapore}
\address{$^{2}$ School of Engineering and Applied Sciences, Harvard University, Cambridge, MA 02138, USA}
\address{$^{3}$ Department of Physics, Harvard University, Cambridge, MA 02138, USA}
\address{$^{4}$ Department of Organismic and Evolutionary Biology, Harvard University, Cambridge, MA 02138, USA
}

\ead{$^{\ast}$eehou@ntu.edu.sg}
\vspace{10pt}
\begin{indented}
\item[]January 2025
\end{indented}

\begin{abstract}
The archetypal instability of a structure is associated with the eponymous Euler beam, modeled as an inextensible curve which exhibits a supercritical bifurcation at a critical compressive load. In contrast, a soft compressible beam is capable of a subcritical instability, a problem that is far less studied, even though it is increasingly relevant in the context of soft materials and structures. Here, we study the stability of a soft extensible elastic beam on an elastic foundation under the action of a compressive axial force, using the Lyapunov-Schmidt reduction method which we corroborate with numerical calculations. Our calculated bifurcation diagram differs from those associated with the classical Euler-Bernoulli beam, and shows two critical loads, $p^\pm_{\text{cr}}(n)$, for each buckling mode $n$. The beam undergoes a supercritical pitchfork bifurcation at $p^+_{\text{cr}}(n)$ for all $n$ and slenderness. 
Due to the elastic foundation, the lower order modes at $p^-_{\text{cr}}(n)$ exhibit subcritical pitchfork bifurcations, and perhaps surprisingly, the first supercritical pitchfork bifurcation point occurs at a higher critical load.
The presence of the foundation makes it harder to buckle the elastic beam, but when it does so,  it tends to buckle into a more undulated shape. 
Overall, we find that an elastic support can lead to a myraid of buckled shapes  for the classical elastica and one can tune the substrate stiffness to control desired buckled modes---an experimentally testable prediction. 

\end{abstract}

%
%
%
%
%

\section{Introduction}

The scientific study of the elastica goes back to the work of James Bernoulli and Leonhard Euler (see \cite{levien2008elastica} for a recent review). The analysis of the loss of equilibrium associated with the loss of stability (bifurcation) of an axially loaded inextensible elastic filament and the nonlinear development of the resulting shapes was also the first attempt to understand the finite amplitude deformations of elastic structures, summarized for example in books on elasticity~\cite{love2013treatise}, stability theory~\cite{antmanbook} and pattern formation~\cite{hoyle2006pattern}.  While the original impetus for the development of the subject was associated with important applications to civil and mechanical engineering in the context of structural stability and its loss~\cite{todhunter2014history, zaccaria2011structures}, today it is just as relevant to a range of questions in the stability of soft systems. This often requires modifications to the classical problem of the elastica to include understanding the importance of shear deformations, extensional effects and an elastic supporting foundation~\cite{bigoni2012nonlinear}, and lead to a plethora of novel phenomena of relevance to the study of soft matter systems~\cite{van2024soft, sahin2012physical}.
Here, we consider one specific modification that has seemingly not been addressed carefully, an extensible Euler-Bernoulli beam on a Winkler foundation under an axial compressive load $P$. 
Our research builds on earlier work that studies the bifurcation point and buckled shapes of an unsupported extensible elastica \cite{magnusson2001behaviour}. 
Similarly, we set boundary conditions of a pinned-pinned beam where the transverse displacements and moments are zero at both ends of the beam and the longitudinal displacement is zero at the left end.
In Sec.~\ref{sec:model}, we derive the system of equations of our elastic models from energy minimization, resulting in a set of two coupled 2nd order nonlinear ordinary differential equations. We then study the linearized problem in order to understand the stability of equilibrium shapes. 
We calculate that the eigenvalues of the equilibrium equations to determine the bifurcation points in Sec.~\ref{sec:stability}. 
In Sec.~\ref{sec:bifurcation}, we consider the bifurcation diagram of the softly supported elastica and plotted some characteristic deformed shapes. 
Using the Liapunov-Schmidt reduction method, we study the normal forms of the resultant scalar equation \cite{golubitsky2012singularities} at each bifurcating points in Sec.~\ref{sec:LS}.
Due to the presence of the elastic foundation, the critical buckling load increases and the straight beam remains stable for a larger range of compressive loads. 
Upon buckling, the beam tends to form a more undulated shape as compared to the case without the Winkler foundation \cite{magnusson2001behaviour}. 
We conclude with final observations and a summary in Sec.~\ref{sec:conclusions}. 
The analytical results agree with our numerical solutions and reveal how subcriticality arising from an elastic foundation  can promote buckling into higher order modes.

\section{Mathematical model} 
\label{sec:model}

\subsection{Description of problem}
Our analysis of the buckling states of an axially loaded, elastically supported, extensible plane filament is based on the Euler-Bernoulli theory of elastic beam \cite{magnusson2001behaviour,reissner1972one, landau2012theory, keener2018principles, bigoni2012nonlinear}, which assumes that plane cross sections remain plane and perpendicular to the neutral axis after deformation, i.e., there is no shear. This assumption can be justified via an asymptotic expansion in the aspect ratio that leads to reduced-order theories from 3-d elasticity~\cite{ciarlet1988three}.

\begin{figure}[htb!]
\centering
\includegraphics[width= 0.6\textwidth]{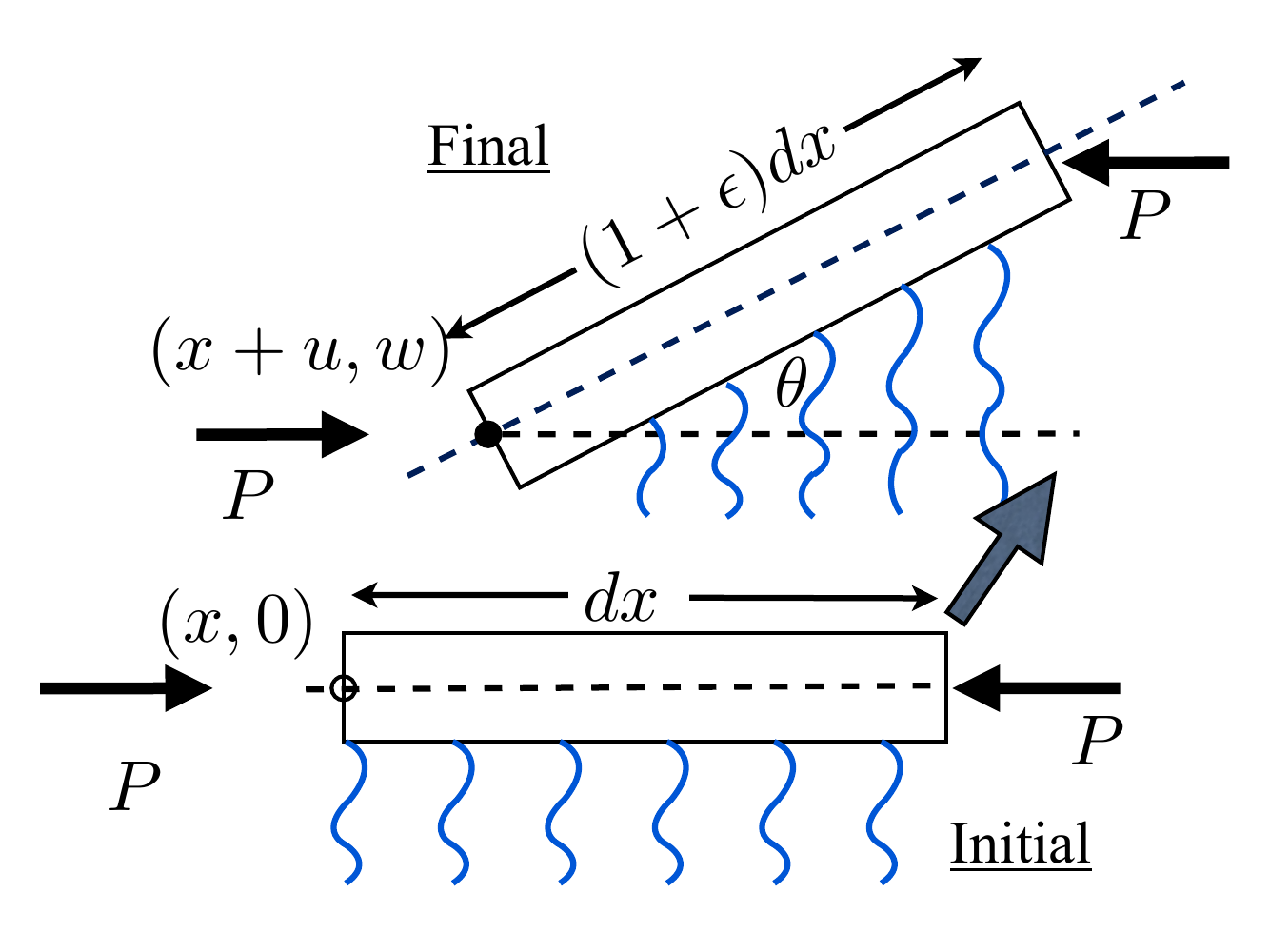}
\caption{Deformation of an elastic extensible beam on a soft substrate. The point $(x, 0)$ moved to $(x+u(x), w(x))$ after deformation where $w(x)$ is the vertical displacement and $u(x)$ is the horizontal displacement. The length of the segment $dx$ is stretched to $ds = (1+\epsilon)dx$, where $\epsilon =  \sqrt{(1+ u_x)^2 + w_x^2 } -1$. Blue curly lines denote the elastic Hookean springs. }
\label{setup}
\end{figure}

In Fig.~\ref{setup}, we show a rod of length $\ell$ initially along the horizontal axis connected to a supporting foundation  of elastic Hookean springs. The material that is initially at $(x, 0)$ will be deformed to $(x', y') = (x + u(x), w(x))$. A small element along the rod $dx$ becomes $ds = (1+ \epsilon)dx$ after deformation where  $\epsilon$ is the midline strain. It experiences a Hookean restoring force in the vertical direction proportional to $w$, i.e., Winkler model \cite{winkler1867lehre, eisenberger1985exact, thambiratnam1996dynamic}. \cite{nicolau1982compressible}. Like in earlier work, we assume that during
deformation, the foundation springs shift freely in the $x$-direction so as to remain vertical and exert forces in the $y$ direction and that they remain attached to the member at fixed points along the center line \cite{nicolau1982compressible}.

Let $\theta$ be the angle that the tangent to the midline makes with the horizontal axis and denote differentiations by subscript $x$, i.e., $A_x \equiv d A/d x$. From geometric consideration, we find that 
\begin{equation}
\cos\theta(x) = \frac{1 + u_x}{1+\epsilon} , \quad \sin\theta(x) = \frac{w_x}{1+\epsilon},
\label{geom}
\end{equation} 
where
\begin{equation}
\frac{ds}{dx} =  1+ \epsilon = \sqrt{(1+ u_x)^2 + w_x^2 }.
\end{equation}
The curvature is 
\begin{equation}
\kappa = \frac{d\theta}{ds} =  \frac{d\theta}{dx}  \frac{dx}{ds} = \frac{\theta_x}{1+\epsilon}.
\end{equation}

\subsection{Variational approach}

The total energy of the system is given by  
\be
U =  \int^\ell_0 dx  \left( \frac{E I}{2}  \theta_x^2 + \frac{EA}{2}  \epsilon^2 + \frac{K}{2} w^2 \right) + P \Delta u,
\label{eq:PE}
\ee
where $E$ is the Young's modulus, $I$ is the moment of inertia, $A$ is the cross-sectional area, $K$ is the Hookean spring constant, and $P$ is the applied axial force. The term with $EI\theta_x^2$ is the elastic strain contribution due to bending, the term with $EA\epsilon^2$ is the elastic strain contribution due to stretching, and the term with $Kw^2$ represents the elastic spring support energy per unit length, and $P \Delta u$ represents the work done by the applied axial force $P$. Since 
\be
\Delta u = u(\ell) - u(0) = \int^\ell_0 dx [(1+\epsilon) \cos \theta - 1],
\ee
we find that we can rewrite Eq.~(\ref{eq:PE}) as
\be
U =  \int^\ell_0 dx  \left[ \frac{1}{2} E I \theta_x^2 + \frac{1}{2} EA \epsilon^2 + \frac{1}{2} Kw^2  
 + P ((1+\epsilon) \cos \theta - 1) \right]. 
 \label{eq:U}  
\ee
Requiring the first variation of the potential energy to vanish, we find that 
\be
\delta U = \int^\ell_0 dx \left\{ EI \theta_{x}\delta\theta_x + EA\epsilon\delta\epsilon + Kw\delta w + P\cos\theta \delta\epsilon - (1 + \epsilon) P \sin\theta \delta \theta \right\} = 0.
\label{eq:deltaU}
\ee
Integrating by parts, the integral becomes
\begin{align}
\delta U &= \int^\ell_0 dx \left\{ P \cos\theta - K\eta \sin\theta + EA \epsilon \right\} \delta \epsilon + [EI \theta_x \delta \theta]^\ell_0 +  [K\eta \delta w]^\ell_0	\nonumber \\
& \quad + \int^\ell_0 dx \{ -EI \theta_{xx} - (1 + \epsilon) P \sin\theta  - K\eta (1+\epsilon) \cos\theta \} \delta \theta = 0,
\end{align}
where we have introduced a new variable $\eta_{x}=w$ and made use of
\be
\delta w_x =  \sin\theta \delta \epsilon + (1+\epsilon)\cos\theta \delta\theta.
\ee
The two boundary terms vanish due to the boundary conditions (for a pinned-pinned beam):
\be
\theta_x(0) = \theta_x(\ell) = w(0) = w(\ell) = 0.
\label{BC}
\ee
Additionally, we set $u(0)=0$. The is a boundary value problem with boundary conditions analogous to earlier work by Magnusson et al. \cite{magnusson2001behaviour}. 
This differs from work by Nicolau and Huddleston who used asymmetric boundary conditions leading to asymmetrical shapes in its post-buckled state \cite{nicolau1982compressible}. 
Since the variations $\delta \epsilon$, $\delta \theta$ are arbitrary, the terms in each parenthesis must necessarily vanish separately leading to
\be
EA \epsilon =  K\eta \sin\theta -P \cos\theta 
\label{eq:strain}
\ee
and
\be
EI \theta_{xx} = - (1 + \epsilon) P \sin\theta  - K\eta (1+\epsilon) \cos\theta.
\ee
We can then eliminate $\epsilon$ to obtain two coupled 2nd order ordinary differential equations (ODEs):
\be
\theta_{xx} = -\alpha P (1 - \beta P \cos \theta + \beta K \eta \sin\theta)\sin\theta -\alpha K \eta(1 - \beta P \cos \theta + \beta K \eta \sin\theta)\cos\theta \label{eq:m1}
\ee
and
\be
\eta_{xx} = (1 - \beta P \cos\theta + \beta K \eta \sin\theta)\sin\theta, \label{eq:m2}
\ee
where $\alpha = (EI)^{-1}$ and $\beta = (EA)^{-1}$. These are the equilibrium equations for an axially loaded beam on a flexible spring support. 

\subsection{$\mathbb{Z}_2$-symmetry}

 Notice that this problem has reflectional symmetry or $\mathbb{Z}_2$-symmetry: $(\theta, \eta) \to (-\theta, -\eta)$. The trivial solution is $(\theta, \eta) = (0, 0)$, i.e., the fundamental solution. We are interested in the bifurcation from the fundamental solution. Let us expand the ODEs in a Taylor series as follows:
\begin{align}
\theta_{xx} &=  - \alpha K (1 - \beta P)\eta - \alpha P (1 - \beta P) \theta-\alpha \beta K^2 \eta^2\theta \nonumber \\
& + \frac{1}{2} \alpha K (1 - 4\beta P)\eta\theta^2 + \frac{1}{6}\alpha P(1 - 4\beta P)\theta^3 + ... \label{eq:duff}\\
\eta_{xx} &= (1 - \beta P)\theta + \beta K \eta\theta^2 - \frac{1}{6} (1 - 4\beta P) \theta^3 + ... \label{eq:eta}
\end{align}
Note that when we set $K=0$ into Eq.~(\ref{eq:duff}), we recover the well-known Duffing's equation \cite{keener2018principles}. 

\section{Stability analysis}
\label{sec:stability}

Let us rewrite the system of ODE Eqs.~(\ref{eq:duff}) and (\ref{eq:eta}) in an abstract form more suitable for analysis \cite{golubitsky2012singularities}. Let
\begin{equation}
\Phi(\theta, \eta) = L \left(\begin{array}{c}
\theta \\ 
\eta
\end{array} \right) + N(\theta, \eta) = 0,
\label{eq:ME}
\end{equation}
where the linear part $L \left(\begin{array}{c}
\theta \\ 
\eta
\end{array} \right)$ is given by
\begin{equation}
\frac{d^2}{d x^2} \left(\begin{array}{c}
\theta \\ 
\eta
\end{array} \right) + \begin{pmatrix}
\alpha P(1-\beta P) & \alpha K(1-\beta P) \\
-(1-\beta P)&0 \end{pmatrix}\left(\begin{array}{c}
\theta \\
\eta
\end{array} \right),
\label{eq:linear}
\end{equation}
and the nonlinear part is given by
\begin{equation}
N(\theta, \eta) = \left(\begin{array}{c}
\alpha \beta K^2 \eta^2\theta - \frac{1}{2} \alpha K (1 - 4\beta P)\eta\theta^2 - \frac{1}{6}\alpha P(1 - 4\beta P)\theta^3 \\ 
-\beta K \eta\theta^2 + \frac{1}{6} (1 - 4\beta P) \theta^3
\end{array} \right).
\end{equation}

In this form, $\Phi(\theta, \eta; P, K, \alpha, \beta)  : \mathscr{X} \times \mathbb{R}^{4} \to \mathscr{Y}$ is a mapping between Banach spaces defined as follows. The domain $\mathscr{X}$ is the space of all real-valued, twice continuously differentiable vector function ${\bm q}(x) = (\theta(x), \eta(x))^{T}$ defined on $x \in [0, \ell]$ that has vanishing first derivatives at the boundaries, i.e. ${\bm q}_x(0) = {\bm q}_x(\ell) = (0, 0)^T$. $\mathbb{R}^{4}$ refers to the space spanned by the four real variables $\{P, K, \alpha, \beta\}$. The range $\mathscr{Y}$ is the space of continuous vector function defined on $x \in [0, \ell]$. Observe that $\Phi(0, 0; P, K, \alpha, \beta) = 0$ for all $P$, $K$. In other words, the undeformed configuration satisfies the equilibrium equations for any external load $P$ and spring constant $K$. 


\subsection{Characteristic buckled load and wavelength of buckled beam}
Firstly, let us consider the linear part of the coupled equations, which read
\begin{align}
\theta_{xx} + \alpha K (1 - \beta P)\eta + \alpha P (1 - \beta P) \theta = 0, \label{eq:scaling1}\\
\eta_{xx} - (1 - \beta P)\theta =0.\label{eq:scaling2}
\end{align}
We can differentiate Eq.~(\ref{eq:scaling1}) twice w.r.t. $x$ and substitute Eq.~(\ref{eq:scaling2}) to get
\begin{equation}
\theta_{xxxx} + \alpha P (1 - \beta P) \theta_{xx} + \alpha K (1 - \beta P)^2\theta =0.
\label{eq:scaling3}
\end{equation}
We consider the ansatz $\theta = e^{\Omega x}$, which yields
\begin{equation}
\Omega^2 = \frac{-\alpha P (1 - \beta P)\pm \sqrt{\left(\alpha P (1 - \beta P)\right)^2-4\alpha K (1 - \beta P)^2}}{2}
\end{equation}
We observe that $\Omega$ is purely imaginary when the argument of the inner square root is positive, which leads to
\begin{equation}
P > P_* = 2\sqrt{\frac{K}{\alpha}},
\end{equation}
where $P_*$ is the characteristic buckled load. When this condition is satisfied, the elastica has periodic buckled solutions. 

Let us define the slenderness of the beam as $\lambda = \ell \sqrt{EA/(EI)}$ \cite{magnusson2001behaviour}. Consider the limit where the beam is long and inextensible, i.e., high slenderness and $\beta P_* \ll 1$. At the critical load, we find that Eq.~(\ref{eq:scaling3}) becomes
\begin{equation}
\theta_{xxxx} + \alpha P_* (1 - \beta P_*) \theta_{xx} + \alpha K (1 - 2 \beta P_*)\theta \simeq 0,
\end{equation}
which has solution
\begin{equation}
\theta(x) = A \cos \left(\frac{x}{\xi}\right), \quad \xi =\left(\frac{EI}{K}\right)^{1/4}.
\label{eq:scaling4}
\end{equation}
Thus, we find that the buckled beam has characteristic wavelength $\xi$, which is a known result from literature \cite{landau2012theory, michaels2019geometric}.

\subsection{Critical buckling loads}

We are interested in bifurcation from the fundamental solution, $(\theta, \eta) = (0, 0)$. In this case, it suffices to consider the linearized problem as described by Eq.~(\ref{eq:linear}). We claim that all eigenfunctions of linear part $L$ have the form
\begin{equation}
{\bm v}_n =  \left(\begin{array}{c}
\theta_n \\ 
\eta_n
\end{array} \right) = \cos\left(\frac{n\pi x}{\ell}\right) \left(\begin{array}{c}
c_1 \\ 
c_2
\end{array} \right),
\label{eigen}
\end{equation}
where $n$ is a positive integer. Observe that the two parameters $c_1$ and $c_2$ span a two-dimensional subspace of functions which is invariant for $L$; additionally, they satisfy the boundary conditions given by Eq.~(\ref{BC}). For each $n$, there exist two linearly independent eigenfunctions of $L$. By Fourier analysis, 
\begin{equation}
\left\{ \cos\left(\frac{n\pi x}{\ell}\right): n = 1,2,3,\dots\right\}
\label{eq:periodic}
\end{equation}
is a complete set of scalar functions. Arguing component-wise, it follows that Eq.~(\ref{eigen}) provides a complete set of vector functions. Thus, we can restrict $L$ to this two-dimensional subspace, namely $L_2$, which gives the matrix
\begin{equation}
L_2=\begin{pmatrix}
-\gamma^2 + \alpha P(1-\beta P) & \alpha K(1-\beta P) \\
-(1-\beta P)& -\gamma^2 \end{pmatrix},
\label{eq:L2}
\end{equation}
where $\gamma = n\pi/\ell$. 
It is useful to define the Euler load, $P_E =  \pi^2 EI/\ell^2$, normalized pressure $p = P/P_E$, and normalized spring constant $\zeta=K/P_E$. 
\begin{equation}
L_2 =\begin{pmatrix}
-n^2 \left(\frac{\pi}{\ell}\right)^2 + \left(\frac{\pi}{\ell}\right)^2 p \left(1-  \left(\frac{\pi}{\lambda}\right)^2 p\right) &  \left(\frac{\pi}{\ell}\right)^2 \zeta \left(1- \left(\frac{\pi}{\lambda}\right)^2 p\right) \\
-\left(1-\left(\frac{\pi}{\lambda}\right)^2 p\right)& -n^2 \left(\frac{\pi}{\ell}\right)^2
\end{pmatrix}. 
\label{eq:L2A}
\end{equation}

\begin{figure}[htb!]
\centering
\includegraphics[width= \textwidth]{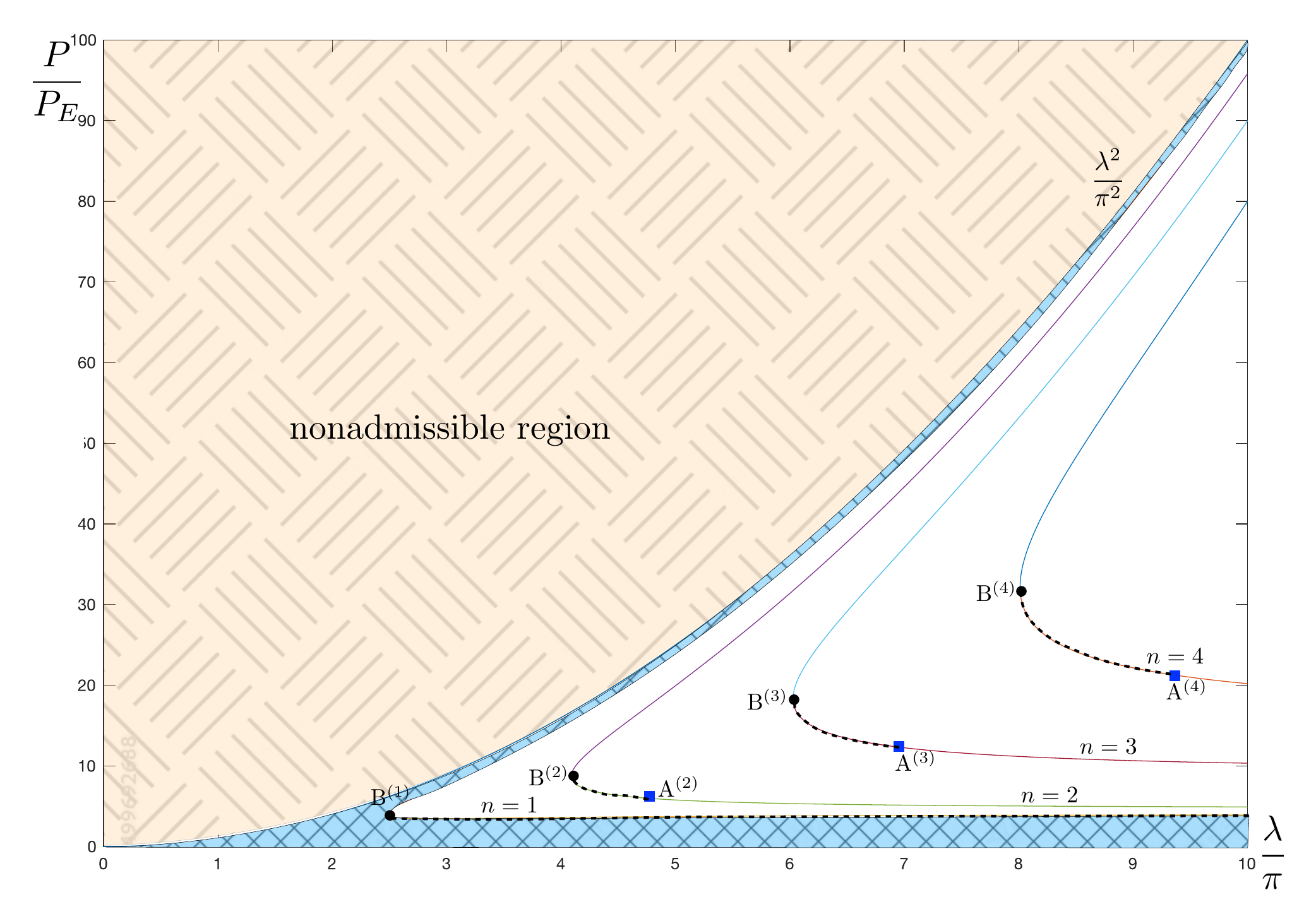}
\caption{Plot of normalized buckling load, $p=P/P_E$, at the fundamental path (blue hatched region) as a function of slenderness, $\lambda/\pi$, for the first four modes and $\zeta=3$. For $n=1$ mode, the beam undergoes supercritical/subcritical pitchfork bifurcation along the upper/lower branch starting from $B^{(1)}$. For higher order modes, i.e., $n \ge 2$, the beam undergoes supercritical pitchfork bifurcation along each solid branch and subcritical pitchfork bifurcation in the intervals $A^{(n)}-B^{(n)}$. 
The upper left triangular region (yellow hatched region), $p>(\lambda/\pi)^2$, denotes the nonadmissible region. }
\label{pic:phase}
\end{figure}

The linear map $L_2$ is invertible unless the determinant vanishes in which case, the implicit function theorem fails and we have a bifurcation. 
The determinant is given by
\begin{equation}
\text{det}(L_2) = n^2 \left(\frac{\pi}{\ell}\right)^4 \left[ n^2   - p \left(1 - \left(\frac{\pi}{\lambda}\right)^2 p\right) \right] + \left(\frac{\pi}{\ell}\right)^2 \zeta \left[1 - \left(\frac{\pi}{\lambda}\right)^2 p\right]^2,
\label{eq:det}
\end{equation}
which is a quadratic equation in $p$. 
The determinant vanishes when the axial load $p$ is equal to
\be
p_{\text{cr}}^{\pm}(n) = \frac{\lambda^2}{2\pi^2} \left\{   \frac{ 2\left(\frac{\pi}{\lambda}\right)^2 \zeta +  \left(\frac{\pi}{\ell}\right)^2 n^2 \pm  \left(\frac{\pi}{\ell}\right)^2 n^2 \sqrt{1 - \frac{4\ell^2}{\lambda^2} \left[\left(\frac{\pi}{\lambda}\right)^2 \zeta + \left(\frac{\pi}{\ell}\right)^2 n^2\right]}}{ \left(\frac{\pi}{\lambda}\right)^2\zeta +  \left(\frac{\pi}{\ell}\right)^2n^2}\right\}, 
\label{eq:proots}
\ee
for $n = 1,2,3,\dots$. Thus, we find that there are two critical loads for each buckling mode $n$. 

For the special case of $\zeta=0$, we find that the Eq.~(\ref{eq:proots}) simplifies to
\be
p_{\text{cr}}^{\pm}(n, \zeta=0) = \frac{\lambda^2}{2\pi^2} \left\{  1  \pm   \sqrt{1 - \frac{4\pi^2 n^2}{\lambda^2}} \right\}, 
\ee
which agrees with earlier work by Magnusson et al. \cite{magnusson2001behaviour}.

Given a specific slenderness $\lambda$ and normalized spring constant $\zeta$, we find that there are two critical buckling loads for each buckling eigenmode and a finite number of buckling modes, unlike the inextensible beam, as illustrated in Fig.~\ref{pic:phase}. For example, for $\lambda = 7.5\pi$ and $\zeta=3$, there are three eigenmodes and six bifurcation points (i.e., $p_{\text{cr}}^{\pm}(n)$, $n=1,2$ and 3). 
Notice that $p_{\text{cr}}^{\pm}(n)$ is real-valued only when
\begin{equation}
1 - \frac{4\ell^2}{\lambda^2} \left[\left(\frac{\pi}{\lambda}\right)^2 \zeta + \left(\frac{\pi}{\ell}\right)^2 n^2\right] \ge 0,
\end{equation}
which implies
\be
\lambda^2 \ge \lambda_0^2(n) = 2\pi^2n^2\left( 1 +  \sqrt{1 + \left(\frac{\ell}{\pi}\right)^2 \frac{\zeta}{n^4}}\,\right),
\ee
where only the positive root is kept since $\lambda^2$ is always nonnegative. When $\lambda^2 <  \lambda^2_0(1)$, $p^\pm_{\text{cr}}(1)$ is not real, implying that bifurcation is not possible and the fundamental mode, $(\theta, \eta) = (0, 0)$, is the only solution. In this case, the compression of the beam is sufficient to accommodate the increased in axial load so that buckling is circumvented. When $\lambda^2 = \lambda^2_0(n)$, we have a double root, $p_{\text{cr}}^{+}(n) = p_{\text{cr}}^{-}(n) = p_B{(n)}$, and there is only one bifurcation axial load where
\be
p_B(n) =  \frac{\lambda^2}{2\pi^2} \left\{   \frac{ 2\left(\frac{\pi}{\lambda}\right)^2 \zeta +  \left(\frac{\pi}{\ell}\right)^2 n^2}{ \left(\frac{\pi}{\lambda}\right)^2\zeta +  \left(\frac{\pi}{\ell}\right)^2n^2}\right\}.
\ee
These special points, i.e., $(\lambda_0(n), p_B{(n)})$, are labeled as $B^{(n)}$ for $n=1,2,3,\dots$ in Fig.~\ref{pic:phase}.


Similar to earlier works, we assume that our material model for the most part behaves in a Hookean manner, i.e., linear, until the spring is compressed to its end position at which point it becomes infinitely stiff. This simplified constitutive linear relation between resultant force and strain results in an admissible region given by $\epsilon > -1$ \cite{antmanbook, magnusson2001behaviour}. Together with Eq.~(\ref{eq:strain}), we find that
\be
\epsilon = \left(\frac{\pi}{\lambda}\right)^2 \zeta \eta \sin\theta -  \left(\frac{\pi}{\lambda}\right)^2 p \cos\theta > -1. 
\ee
Along the fundamental path, i.e. $(\theta, \eta) = (0, 0)$, the admissible region is given by the condition
\be
p=\frac{P}{P_E} < \left(\frac{\lambda}{\pi}\right)^2.
\label{eq:admiss}
\ee
The admissible region occupies the lower right triangular region of the $P/P_E$ versus $\lambda/\pi$ phase space as shown in Fig.~\ref{pic:phase}.

\section{Bifurcation diagrams}
\label{sec:bifurcation}

\subsection{Inextensible Elastica}

In the case of the classical inextensible elastica, the coupled ODEs (Eqs.~(\ref{eq:m1}) and (\ref{eq:m2})) become
\be
\theta_{xx} + \left(\frac{\pi}{\ell}\right)^2 p \sin\theta + \left(\frac{\pi}{\ell}\right)^2 \zeta\, \eta \cos\theta = 0
\ee
and
\be
\eta_{xx}  - \sin\theta = 0.
\ee
The $L_2$ matrix (Eq.~(\ref{eq:L2A})) becomes
\begin{equation}
L_2^{\text{in}}{\bm v}_n =\begin{pmatrix}
-n^2 \left(\frac{\pi}{\ell}\right)^2 + \left(\frac{\pi}{\ell}\right)^2 p  &  \left(\frac{\pi}{\ell}\right)^2 \zeta  \\
-1 & -n^2 \left(\frac{\pi}{\ell}\right)^2
\end{pmatrix} {\bm v}_n = {\bm 0}.
\end{equation}
The superscript ``in" denotes the inextensible case. In this case, the determinant is given by
\begin{equation}
\text{det}(L_2^{\text{in}}) =  \left(\frac{\pi}{\ell}\right)^4 n^2 \left( n^2   - p \right)+ \left(\frac{\pi}{\ell}\right)^2 \zeta,
\end{equation}
which vanishes when the axial load is equal to
\be
p_{\text{cr}}^{\text{in}}(n)= n^2 + \left(\frac{\ell}{\pi}\right)^2\frac{\zeta}{n^2}.
\label{eq:pinext}
\ee
Thus, there is one critical load for each eigenmode $n$ for the inextensible elastica on soft support and theoretically, an infinite number of buckling modes. 
In this simple case, we find that the presence of the soft spring support provides a correction term to the critical axial load that is proportional to the spring constant.

Since our solution is periodic as evident from Eq.~(\ref{eq:periodic}), we can identify $\ell/(n\pi)$ as a characteristic wavelength $\xi$. If we set $\ell/(n\pi) \sim \xi = 1/(\alpha K)^{1/4}$ and substitute into Eq.~(\ref{eq:pinext}), we find that the characteristic buckled load is $P_{\text{cr}}^{\text{in}} \sim \xi^2 \zeta \sim \sqrt{K/\alpha}$, which is consistent with what we found earlier. 
For the case $\zeta = 0$, we find that $p_{\text{cr}}^{\text{in}}(n) = n^2$, i.e., $P_{\text{cr}}^{\text{in}} = n^2 P_E$, which agrees with earlier work by Magnusson et al. \cite{magnusson2001behaviour}.

\begin{figure}[htb!]
\centering
\includegraphics[width= \textwidth]{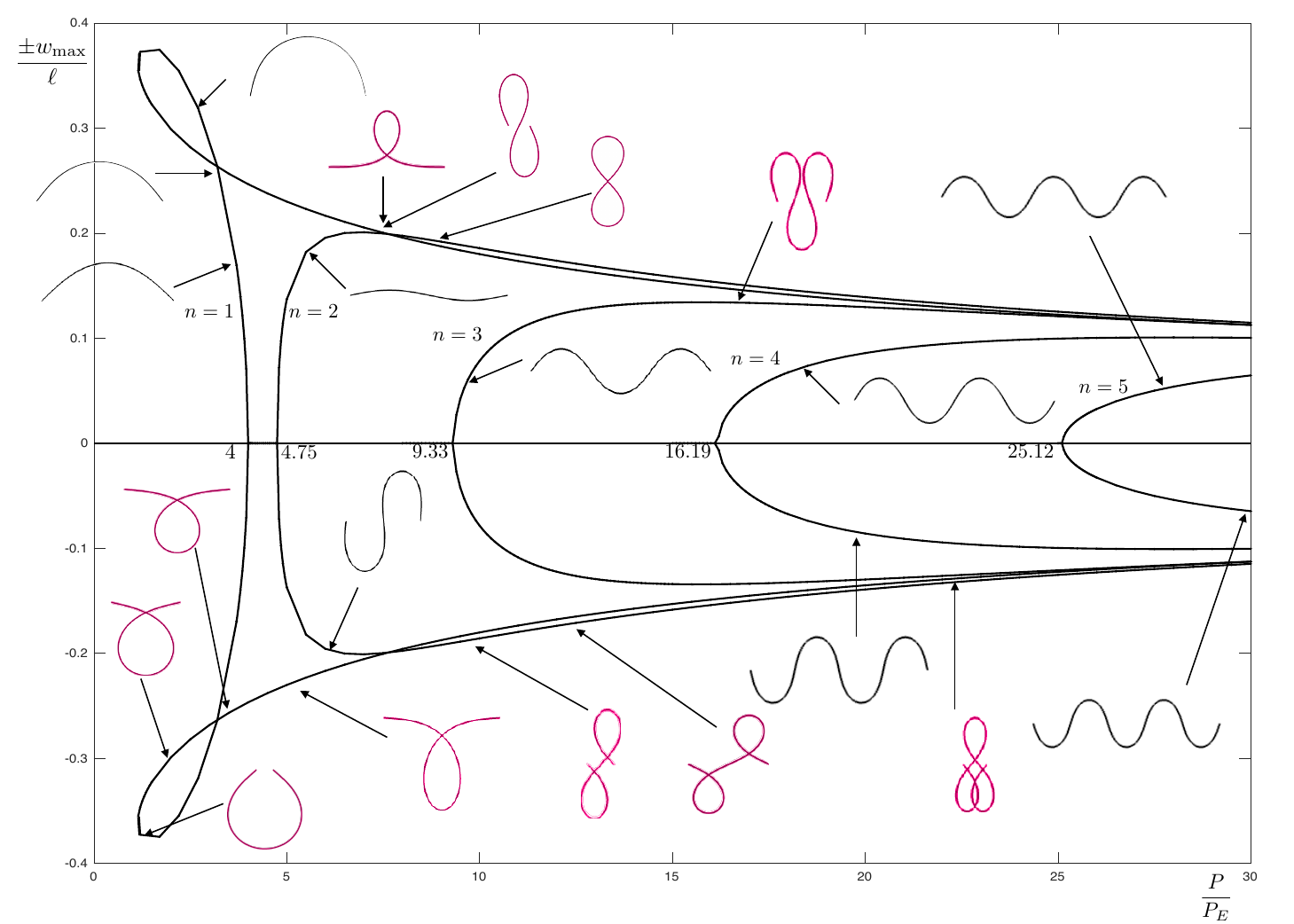}
\caption{Bifurcating diagram $(P/P_E, \pm w_{\text{max}}/\ell)$ of the softly supported inextensible elastica for the first five modes for $\ell=\pi$ and $\zeta=3$. Some characteristic deformed shapes of the beam are shown along different bifurcation branch. Self-intersecting shapes are colored in magenta. }
\label{pic:inext}
\end{figure}

For the elastica problem, a natural scalar to measure bifurcation is the maximum value of $w(x)$ over the domain $x\in[0, \ell]$:
\begin{equation}
w_{\text{max}} =\max_{x\in[0, \ell]} |w(x)|.
\end{equation}
The postbuckling behavior for the classical inextensible elastica on a soft support, $\pm w_{\text{max}}/\ell$, as a function of the axial load $P/P_E$ is plotted in Fig.~\ref{pic:inext} for $\ell=\pi$, $\zeta=3$, and $n=1,2,3,4,5$. This is computed in Matlab using method of continuation \cite{trefethen2017exploring}. Our theory imposes no restrictions on the size of the displacements and thus it is possible to analyze the entire range of postbuckling shapes.

Some characteristic shapes of the elastica are illustrated along the different bifurcating solution branches. 
The $n=1$ mode exhibits a subcritical pitchfork bifurcation at $p_{\text{cr}}^{\text{in}}(n=1) = 4$. 
The maximal deflection for $n=1$ branch first increases with decreasing load along the unstable branch from $p = 4$ until around $p \approx 1.15$ and then start to decrease with increasing load. This happens when the two end points pass each other generating a loop, allowing for lower maximal deflection. 
These buckled shapes that self-intersect are colored in magenta in Fig.~\ref{pic:inext}. 
As the pressure is decreased further, this solution branch with loop crosses the state without loop at $p \approx 3.2$. Such intersection is an example of ``crossed'' fold bifurcation \cite{chen2020biocrust}. 

At higher pressure, this bifurcation branch continues to decrease with increasing pressure until at $p \approx 7.5$ where we observe the $n=1$ branch undergoes a transcritical bifurcation into the $n=2$ mode \cite{armitstead1996study}. 
All the higher order bifurcation modes ($n\ge2$) exhibit supercritical pitchfork bifurcation at their respective critical load given by Eq.~(\ref{eq:pinext}). For each mode $n \ge 2$, the maximal deflection $\pm w_{\text{max}}/L$ first increases with increasing load and then decrease at higher loads when the two end-points pass each other, generating $n$ loops. 
As $P/P_E$ increases further, more and more solution branches appear. We shall not attempt to track them graphically. When considering multivalued buckling loads, practical significance should be assigned only to the smallest modes. 
As the $n=1$ mode is subcritical, we find that the buckling mode of practical interest is the $n=2$ buckling mode, which is the first supercritical pitchfork bifurcation. 



\subsection{Extensible elastica}

\begin{figure}[htb!]
\centering
\includegraphics[width= \textwidth]{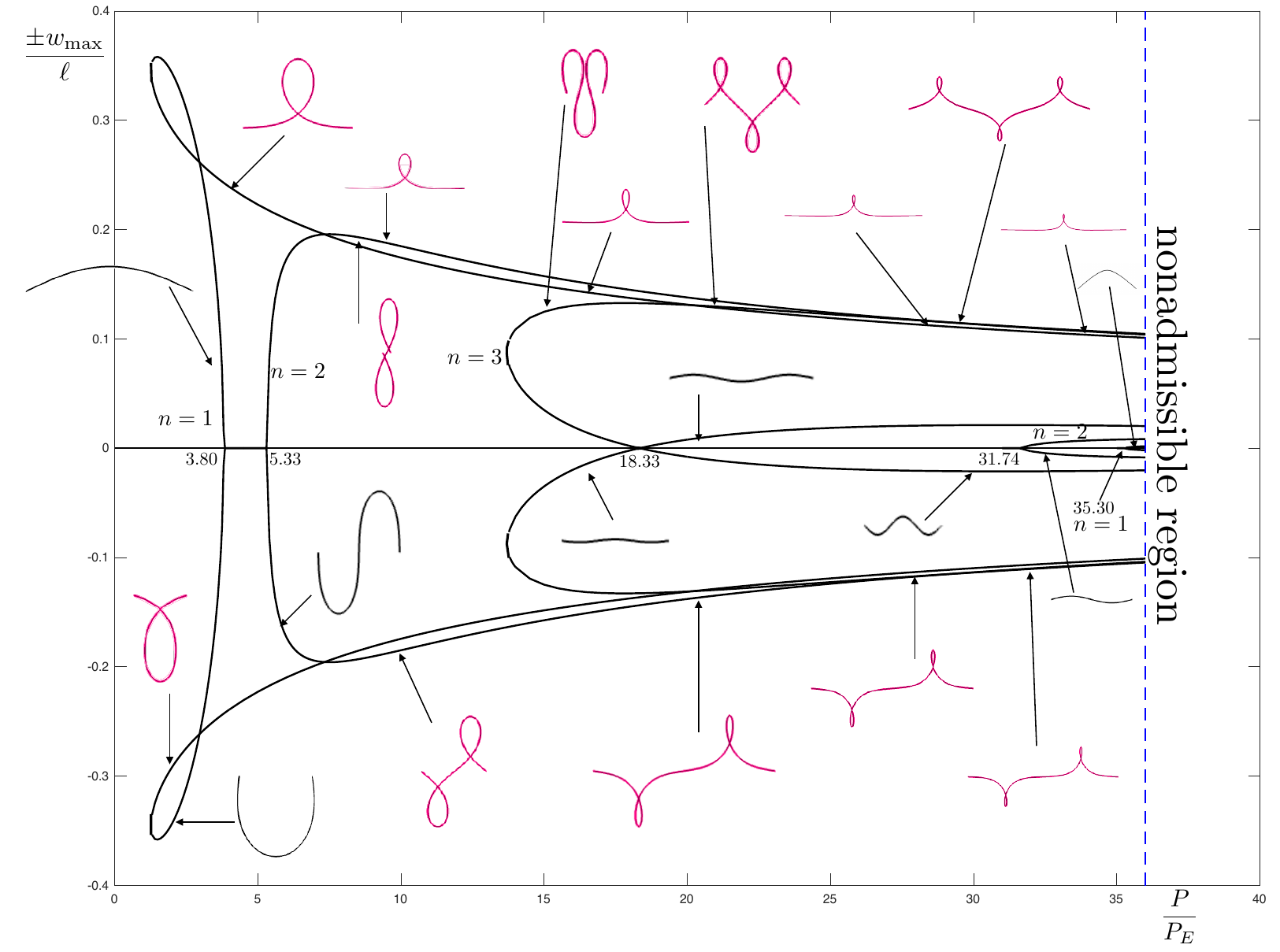}
\caption{Bifurcating diagram $(P/P_E, \pm w_{\text{max}}/\ell)$ of the softly supported extensible elastica for slenderness $\lambda = \lambda_0(3) \approx 6.03 \pi$, $\ell=\pi$, and $\zeta=3$. 
Some characteristic deformed shapes of the beam are shown along each bifurcation branch. 
Self-intersecting shapes are colored in magenta.
}
\label{pic:ext}
\end{figure}

The postbuckling behavior for the classical extensible elastica on a soft support, $\pm w_{\text{max}}/\ell$, as a function of the axial load $P/P_E$ is plotted in Fig.~\ref{pic:ext} for slenderness $\lambda=\lambda_0(3) \approx 6.03 \pi$, $\ell=\pi$, and $\zeta=3$. 
At this slenderness, there are only three bifurcation modes, namely, $n=1,2,$ and $3$, and five critical axial loads at $p \approx 3.80, 5.33, 18.33, 31.74, \text{and } 35.30$ as evident from Figs.~\ref{pic:phase} and \ref{pic:ext} \cite{golubitsky2012singularities}. 
Some characteristic shapes of the elastica are illustrated along the different bifurcating solution branches. 
The admissible region is given by $p \in [0, \lambda_0^2(3)/\pi^2]$ which follows from Eq.~(\ref{eq:admiss}).

Similar to the inextensible case, the $n=1$ mode exhibits a subcritical pitchfork bifurcation at $p_{\text{cr}}(1) \approx 3.80$. 
The maximal deflection for $n=1$ branch first increases with decreasing load along the unstable branch from $p = 4$ until around $p \approx 1.25$ and then start to decrease with increasing load. This happens when the two end points pass each other generating a loop, allowing for lower maximal deflection. As the pressure is decreased further, this solution branch with loop crosses the state without loop at $p \approx 2.95$. At higher pressure, this bifurcation branch continues to decrease with increasing pressure until at $p \approx 7.3$ where we observe the $n=1$ branch undergoes a transcritical bifurcation
into the $n=2$ mode \cite{armitstead1996study}.

The $n=2$ bifurcation branch exhibits a supercritical pitchfork bifurcation at $p\approx 5.33$, which is the effective critical buckling load for the beam. 
The maximal deflection $\pm w_{\text{max}}/L$ first increases with increasing load and then decrease at higher loads when the two end-points pass each other, generating two loops. 
These physically unrealistic buckled shapes are colored in magenta. 
The critical load at $p\approx 18.33$ corresponds to a double root, i.e., point $B^{(3)} = (\lambda_0(3), p_B(3)$), where we see two paths emanating from the singular point. Here, the system exhibits a nondegenerate cubic singularity, i.e., a singularity of codimension five \cite{golubitsky2012singularities}.
At higher values of $p$, we observe two small supercritical pitchfork bifurcation at $p\approx 31.74$ and $35.30$ which corresponds to $p^+_{\text{cr}}(n=2)$ and $p_+^{\text{cr}}(n=1)$ respectively. Beyond $p = \lambda_0^2(3)/\pi^2$ is the nonadmissible region of our extensible elastica as described by Eq.~(\ref{eq:admiss}).



\section{Liapunov-Schmidt reduction}
\label{sec:LS}


In this section, we will determine the types of bifurcation along the $\text{det}(L_2) = 0$ branches in phase space. We will carry out our analysis of Eq.~(\ref{eq:ME}) using Liapunov-Schmidt reduction method which leads to a single scalar equation \cite{golubitsky2012singularities}:
\be
g(s, \delta p) = 0,
\label{eq:g}
\ee
where $s$ is the state variable and $\delta p = p - p_{\text{cr}}$ is the bifurcation parameter measured relative to the critical axial load. As constructed, we see that $g(0,0)=0$. The set of $(s, \delta p)$ is the bifurcation diagram. At a bifurcation point, $g_s(0,0) = 0$ by the implicit function theorem, where the subscript $s$ indicates partial derivatives here. 

First, let us decompose $\mathscr{X}$ and $\mathscr{Y}$ as follows
\begin{align}
\mathscr{X} &= \text{kernel}(L_2) \oplus (\text{kernel}(L_2))^\perp, \\
\mathscr{Y} &= \text{range}(L_2) \oplus (\text{range}(L_2))^\perp.
\end{align}
The one-dimensional kernel of $L_2$ is spanned by $m_1(x)$, where
\be
m_1(x)  = \cos\left(\frac{n\pi x}{\ell}\right) \left(\begin{array}{c}
\gamma^2 \\ 
-(1-\beta P_{\pm})
\end{array} \right)= c(x) \left(\begin{array}{c} 
a_1 \\ 
a_2
\end{array} \right),
\ee
where $c(x) = \cos (n\pi x/\ell)$ and 
$P_\pm=p_{\pm}P_E$ are the solutions to $\text{det}(L_2)=0$ as given by Eq.~(\ref{eq:proots}). 

By the Fredholm alternative, we have 
\begin{equation}
(\text{range}(L_2))^\perp = \text{kernel}(L_2^*).
\end{equation}
The one-dimensional kernel of
\begin{equation}
L_2^* = \begin{pmatrix}
-\gamma^2 + \alpha P(1-\beta P) & -(1-\beta P) \\
\alpha K(1-\beta P) & -\gamma^2 \end{pmatrix}
\end{equation}
is spanned by $n_1(x)$, where
\be
n_1(x) = \cos\left(\frac{n\pi x}{\ell}\right) \left(\begin{array}{c}
\gamma^2 \\ 
\alpha K(1-\beta P_{\pm})
\end{array} \right) = c(x) \left(\begin{array}{c}
b_1 \\ 
b_2
\end{array} \right).
\ee
We find that $L_2$ is Fredholm with index 0 since dim$(\text{kernel}(L_2))$ = codim$(\text{range}(L_2)) = 1$. 

Using $m_1(x)$ as a basis for $\text{kernel}(L_2)$ and $n_1(x)$ as a basis for $(\text{range}(L_2))^\perp$, we can formally calculate the reduced scalar equation given by Eq.~(\ref{eq:g}). 
To each solution $(s, \delta p)$ of the reduced equation, there corresponds a solution of the full equation of the form
\be
\left(\begin{array}{c}
\theta \\ 
\eta
\end{array} \right) = s \cos\left(\frac{n\pi x}{\ell}\right) \left(\begin{array}{c}
\gamma^2 \\ 
-(1-\beta P_{\pm})
\end{array} \right) + O(s^2).
\ee
In other words, $s$ parameterizes $\text{kernel}(L_2)$, i.e., $s\,m_1(x)$.  Since our problem has the trivial solution $\theta = \eta = 0$, from which it follows that $g(0, \delta p) \equiv 0$. Thus, at the singularity at the origin, we have
\begin{equation}
g = g_s = g_{\delta p} = g_{\delta p\delta p} = 0.
\label{con1}
\end{equation}
In order to find the normal form of the scalar function $g(s, p)$, we need to calculate the different derivatives of $g$ with respect to both $s$ and $p$. In the present case, $\Phi$ is odd with respect to $s$, i.e.,
\be
\Phi(-s, \delta p)= -\Phi(s, \delta p). 
\ee
Therefore, at the trivial point \cite{golubitsky2012singularities}, we have
\begin{equation}
(d\Phi)_{0, \delta p} = L_2, \quad (d^2 \Phi)_{0, \delta p} = 0, \quad \text{and} \quad \Phi_{\delta p} = \frac{\partial \Phi}{\partial \delta p}= 0.
\label{eq:vanish}
\end{equation} 
which implies that 
\begin{equation}
g_{rr} = \langle n_1, d^2\Phi(m_1, m_1) \rangle = 0.
\label{con2}
\end{equation}
The fact that $(d^2 \Phi)_{0, \delta p}$ vanishes simplifies calculations such as $g_{rr}$ tremendously as troublesome terms involving $L_2^{-1}$ drop out.
Due to Eqs.~(\ref{con1}), (\ref{eq:vanish}), and (\ref{con2}), the simplest nontrivial bifurcation we expect in the elastica is the pitchfork bifurcation. We find that
\be
g_{r\delta p} = \langle n_1, d\Phi_{\delta p} \cdot m_1 \rangle
\label{gy}
\ee
and
\be
g_{rrr} = \langle n_1, d^3\Phi(m_1,m_1,m_1)\rangle,
\label{gx}
\ee
where the expression for $g_{rrr}$ is considerably uncomplicated due to the fact that $(d^2 \Phi)_{0, \delta p}$ vanishes. 
Note that the $k$-th order differential of a mapping $\Phi$ at a point $u \in \mathscr{X}$ is given by the following formula \cite{golubitsky2012singularities},
\begin{equation}
(d^k\Phi)_u (v_1, ..., v_k) = \frac{\partial}{\partial t_1} \cdots \frac{\partial}{\partial t_k}\Phi\bigg(u + \sum_{i=1}^k t_i v_i\bigg)\Bigg|_{t=0}.
\end{equation}

\subsection{Inextensible elastica}

We will first turn to the case of the classical inextensible elastica before moving to the extensible case in the next subsection. For the classical inextensible elastica, there are an infinite number of critical buckling loads as given by Eq.~(\ref{eq:pinext}). We find that
\be
m_1^{\text{in}}(x) = \cos\left(\frac{n\pi x}{\ell}\right) \left(\begin{array}{c}
\gamma^2 \\ 
-1
\end{array} \right)
\ee
and
\be
n_1^{\text{in}}(x) = \cos\left(\frac{n\pi x}{\ell}\right) \left(\begin{array}{c}
\gamma^2 \\ 
\alpha K
\end{array} \right).
\ee
Therefore, we find that
\be
d \left(\frac{\partial \Phi}{\partial \delta p}\right)\cdot m_1^{\text{in}} =n^2 \left(\frac{\pi}{\ell}\right)^4 \cos\left(\frac{n\pi x}{\ell}\right) \left(\begin{array}{c}
1 \\ 
0
\end{array} \right)
\ee
and
\be
g_{r\delta p} = n^4 \left(\frac{\pi}{\ell}\right)^6 \int_0^\ell \cos^2 \left(\frac{n\pi x}{\ell}\right) dx.
\ee
Thus, we find that $g_{r\delta p}(0,0)$ is always positive for all $n = 1,2,3, \dots$, i.e., $g_{r\delta p}>0$. 

Likewise, we find that
\be
d^3\Phi(m_1^{\text{in}},m_1^{\text{in}},m_1^{\text{in}}) =\gamma^4 \cos^3\left(\frac{n\pi x}{\ell}\right) \left(\begin{array}{c}
-\alpha \gamma^2 P + 3 \alpha K \\ 
\gamma^2
\end{array} \right)
\ee
and
\be
g_{rrr} = \alpha \gamma^6 P_E \left(- \gamma^2 p+ 4 \zeta \right) \int_0^\ell \cos^4 \left(\frac{n\pi x}{\ell}\right) dx.
\ee
The sign of $g_{rrr}(0,0)$ depends the terms in the parenthesis. On substituting the buckling load for the inextensible elastica, i.e., Eq.~(\ref{eq:pinext}), we find that
\be
\text{sgn}\left(-\gamma^2 p_{\text{cr}}^{\text{in}}+ 4 \zeta\right) = \text{sgn}\left(3\zeta - n^4 \left(\frac{\pi}{\ell}\right)^2\right).
\ee

For our analysis in Fig.~\ref{pic:inext}, i.e., $\ell=\pi$ and $\zeta=3$, so $g_{rrr} > 0$ for $n=1$ and $g_{rrr} < 0$ for $n \ge 2$. In other words, the bifurcation from the fundamental mode is a subcritical pitchfork for only the lowest non-trivial mode $n=1$, with normal form given by $r^3+(\delta p) r = 0$. For all higher order modes, $n \ge 2$, the bifurcation from the fundamental mode is a supercritical pitchfork, with normal form given by $-r^3+(\delta p) r=0$. Thus, the normal forms calculated analytically using the method of Liapunov-Schmidt reduction technique  agrees with our numerical results from the previous section. 

As we increase $\zeta$, we find that more of the lower modes will turn subcritical. The first supercritical pitchfork bifurcation mode $p_{\text{cr}}^{\text{in}}$ is the buckling load of practical interest and it can be tuned by the strength of the spring constant. 
For example, for $\ell=\pi$ and $\zeta=20$, $g_{rrr} > 0$ for $n=1$ and 2 and $g_{rrr} < 0$ for $n \ge 3$, and so the first supercritical pitchfork is $n=3$. 
For $\ell=\pi$ and $\zeta=30$, $g_{rrr} > 0$ for $n=1,2$ and 3 and $g_{rrr} < 0$ for $n \ge 4$, we find that the first supercritical pitchfork is now $n=4$. 
Our analysis here pertains to bifurcation from the fundamental mode $(\theta, \eta) = (0,0)$ only. 
We will not attempt to calculate the post-buckling bifurcation between $n=1$ and $n=2$ mode due to technical difficulties although we know they must exist from numerical computations as evident in Fig.~\ref{pic:inext}.

\subsection{Extensible elastica}

\begin{figure}[htb!]
\centering
\includegraphics[width=\textwidth]{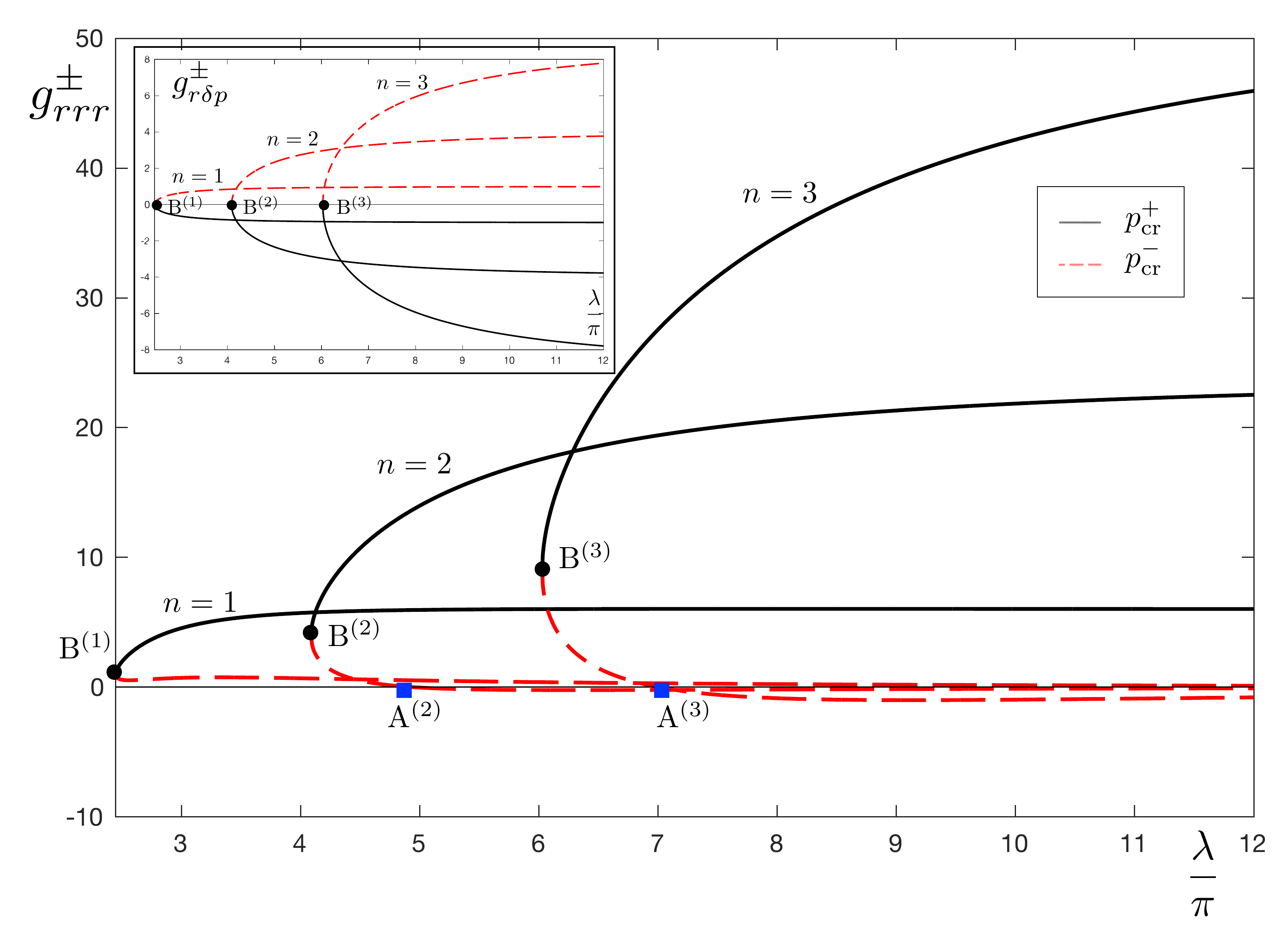}
\caption{Graph of $g_{rrr}^\pm$ (see Eq.~(\ref{eq:grrr})) as a function of slenderness $\lambda/\pi$. The solid lines denote $p^+$; the dashed lines denote $p^-$. 
$g_{rrr}^+$ is always positive while the sign of $g_{rrr}^-$ depends on $\lambda$. 
$g_{rrr}^-$ vanishes when $\lambda = \lambda_c(n)$, labelled by the points $A^{(n)}$ (blue squares). At $\lambda = \lambda_0(n)$, $p^+ (n)= p^-(n)$, and are labelled by the points $B^{(n)}$ (black circles). 
For each mode $n$, $g_{rrr}^- > 0$ for $\lambda_0(n) \le \lambda < \lambda_c(n)$ and $g_{rrr}^- < 0$ when $\lambda > \lambda_c(n)$. 
Inset: Plot of $g_{r\delta p}^\pm$ as a function of $\lambda/\pi$. $g_{r\delta p}^{\pm} = 0$ when $\lambda =\lambda_0(n)$. For each mode $n$, $g_{r\delta p}^+<0$ and $g_{r\delta p}^->0$ for $\lambda > \lambda_0(n)$. }
\label{pic:recognition}
\end{figure}

For the extensible elastica, we find that
\be
d\Phi_{\delta p}\cdot m_1 = d\left(\frac{\partial \Phi}{\partial \delta p}\right)\cdot m_1 = \left(\begin{array}{c}
\alpha(1-2\beta P_{\pm})a_1 - \alpha \beta K a_2\\ 
\beta a_1
\end{array} \right)
\ee
and
\begin{align}
g_{r\delta p}^{\pm} &= \alpha \gamma^2 \left( \gamma^2 (1-2\beta P_{\pm}) + 2\beta K(1-\beta P_{\pm}) \right) \int_0^\ell \cos^2 \left(\frac{n\pi x}{\ell}\right) dx  \nonumber\\
&= \mp \frac{\ell}{2}\alpha \gamma^4 \sqrt{1 - 4\frac{\beta}{\alpha}(\gamma^2+\beta K)} \nonumber\\
&= \mp \frac{\ell}{2}\alpha \gamma^4 \sqrt{1 - \frac{4\ell^2}{\lambda^2} \left[\left(\frac{\pi}{\lambda}\right)^2 \zeta + \left(\frac{\pi}{\ell}\right)^2 n^2\right]}.
\label{eq:gup}
\end{align}
The expression inside the square root of $g_{r\delta p}^{\pm}$ is identical to the one in Eq.~(\ref{eq:proots}). Thus, $g_{r\delta p}$ is real as long as $\lambda \ge \lambda_0(n)$. For critical loads $p^+_{\text{cr}}{(n)}$, we find that $g_{r\delta p}^+ <0$ while for $p^-_{\text{cr}}(n)$, we find that $g_{r\delta p}^- >0$. When $p^+_{\text{cr}}{(n)} = p^-_{\text{cr}}{(n)}$, i.e., $\lambda = \lambda_0(n)$ as indicated by the points $B^{(n)}$ in Fig.~\ref{pic:recognition}, we find that $g_{r\delta p}^{\pm} = 0$.

Similarly, we find that
\begin{equation}
d^3\Phi(m_1, m_1, m_1) = \left(\begin{array}{c}
6 \alpha \beta K^2 a_1 a_2^2 - 3\alpha K(1 - 4\beta P_{\pm})a_1^2 a_2 -\alpha P_{\pm}(1 - 4\beta P_{\pm})a_1^3\\ 
-6\beta K a_1^2 a_2 + (1- 4\beta P_{\pm})a_1^3
\end{array} \right)
\end{equation}
and
\begin{align}
g_{rrr}^\pm &=  B\left(\begin{array}{c}
6 \alpha \beta K^2 a_1 a_2^2 - 3\alpha K(1 - 4\beta P_{\pm})a_1^2 a_2 -\alpha P_{\pm}(1 - 4\beta P_{\pm})a_1^3\\ 
-6\beta K a_1^2 a_2 + (1- 4\beta P_{\pm})a_1^3
\end{array} \right)\cdot \left(\begin{array}{c}
b_1 \\ 
b_2
\end{array} \right) \nonumber \\
 &= - B \gamma^8\, \frac{ 8\beta \gamma^4+32\beta^2\gamma^2K+24\beta^3K^2-3\alpha(\gamma^2 +2 \beta K) \mp 3\gamma^2\sqrt{\alpha(\alpha-4\beta(\gamma^2+\beta K))}}{2\beta (\gamma^2+\beta K)}  \nonumber \\
 &=-B\gamma^8 \frac{\left\{\splitfrac{8 \left(\frac{\pi}{\lambda}\right)^2 \left(\frac{n\pi}{\ell}\right)^4+32 \left(\frac{\pi}{\lambda}\right)^4 \left(\frac{n\pi}{\ell}\right)^2 \zeta  + 24 \left(\frac{\pi}{\lambda}\right)^6\zeta^2 }{- 3 \left(\frac{\pi}{\ell}\right)^2\left(\left(\frac{n\pi}{\ell}\right)^2 + 2\left(\frac{\pi}{\lambda}\right)^2 \zeta\right) \mp 3 \left(\frac{\pi}{\ell}\right)^4 n^2 \sqrt{1 - \frac{4\ell^2}{\lambda^2} \left[\left(\frac{\pi}{\lambda}\right)^2 \zeta + \left(\frac{n\pi}{\ell}\right)^2 \right]} } \right\}}{2 \left(\frac{\pi}{\lambda}\right)^2 \left( \left(\frac{\pi}{\lambda}\right)^2 \zeta +  \left(\frac{n \pi}{\ell}\right)^2\right)}.
 \label{eq:grrr}
\end{align}          
where $B = \int^\ell_0 \cos^4 (n\pi x/\ell) dx = 3\ell/8$. If $\lambda^2 \ge \lambda^2_0(n)$, $g_{rrr}$ is always real. The graphs of $g_{r\delta p}^{\pm}$ and $g_{rrr}^\pm$ as a function of slenderness $\lambda/\pi$ are shown in Fig.~\ref{pic:recognition} for the lowest three modes. 

Unlike the inextensible case where there is one critical load per eigenmode, we have 2 critical loads per eigenmode for the extensible beam. At critical loads $p^+_{\text{cr}}{(n)}$, i.e., positive root in Eq.~(\ref{eq:proots}), we find that $g_{rrr}^+ > 0$ and $g_{r\delta p}^+ < 0$ for all modes $n$ and slenderness $\lambda \ge \lambda_0(n)$.  The bifurcation from the fundamental mode at each buckling loads $p^+_{\text{cr}}{(n)}$ is a supercritical pitchfork, with normal form given by $r^3-(\delta p^+) r = 0$ where $\delta p^+ = p - p^+_{\text{cr}}(n)$ for all $n$. In other words, we have supercritical pitchfork bifurcation for slenderness beyond $B^{(n)}$ along the $p^+_{\text{cr}}{(n)}$ branch for all modes. For $\lambda < \lambda_0(n)$, no critical buckling loads exist on the fundamental path, so there is no bifurcation and the trivial solution $(\theta, \eta)=(0,0)$ is always stable.

The bifurcation at buckling loads $p^-_{\text{cr}}{(n)}$ is more complex: even though $g_{r\delta p}^-$ is always positive, $g_{rrr}^-$ and can be either positive or negative. $g_{rrr}^-=0$ at critical values of slenderness $\lambda_c(n)$ which are labelled as $A^{(n)}$ in Figs.~\ref{pic:phase} and \ref{pic:recognition}. The expression for $\lambda_c(n)$ will not be shown as they are overly complicated and not amenable for analytical calculations. 
For the extensible beam considered in Fig.~\ref{pic:ext}, i.e., $\ell=\pi$, $\zeta=3$, we find that $\lambda_c(2)\approx 4.94$, $\lambda_c(3)\approx 7.00$ and $\lambda_c(3)\approx 9.27$. 
For $n=1$ mode, there is no $\lambda_c(1)$ as $g_{rrr}^-$ is always positive. Since $g_{rrr}^->0$ and $g_{r\delta p}^->0$, the normal form is given by $r^3+(\delta p^-) r = 0$ where $\delta p^- = p - p^-_{\text{cr}}(1)$, i.e., the extensible beam undergoes subcritical pitchfork bifurcation at $p^-_{\text{cr}}(1)$. Thus, we have 
subcritical pitchfork bifurcation along the whole $p^-_{\text{cr}}{(1)}$ branch. 

For $n\ge2$, for slenderness $\lambda_0(n) < \lambda < \lambda_c(n)$, we have $g_{rrr}^->0$ and $g_{r\delta p}^->0$ and we find that the normal form is $r^3+(\delta p^-) r = 0$ where $\delta p^- = p - p^-_{\text{cr}}(n)$, i.e., the beam undergoes subcritical pitchfork bifurcation. For slenderness $\lambda > \lambda_c(n)$, we get $g_{rrr}^->0$ and $g_{r\delta p}^-<0$, so the normal form is $-r^3+(\delta p^-) r = 0$, i.e., the beam undergoes supercritical pitchfork bifurcation. In other words, for $n\ge2$ modes, we have subcritical pitchfork bifurcation for slenderness between $B^{(n)}$ and $A^{(n)}$ along the $p^-_{\text{cr}}{(n)}$ branch and supercritical pitchfork bifurcation for slenderness beyond $A^{(n)}$ along the $p^-_{\text{cr}}{(n)}$ bifurcation branch. Note that for $\lambda = \lambda_c(n)$, $g_{rrr}^-$ vanishes and the normal form of $g(r, \delta p)$ takes on a canonical form given by $\pm r^5 + (\delta p) r = 0$, provided the fifth-order derivative is nonzero \cite{golubitsky2012singularities}.

For an extensible elastica with slenderness $\lambda= \lambda_0(3) \approx 6.03 \pi$, we find the following results: At $p^-_{\text{cr}}(1) \approx 3.80$, the beam undergoes a subcritical pitchfork bifurcation with normal form $r^3+(\delta p^-) r = 0$. At $p^-_{\text{cr}}(2) \approx 5.33$, $p^+_{\text{cr}}(2) \approx 31.74$, and $p^+_{\text{cr}}(1) \approx 35.30$, the beam undergoes a supercritical pitchfork bifurcation with normal form $r^3-(\delta p^\pm) r = 0$. At $p^\pm_{\text{cr}}(1) \approx 18.33$, we have a double root in $p$ for the third mode at $B^{(3)}$, and the normal form is of the form $-r^3 + (\delta p)^2 r =0$, which is a nondegenerate cubic singularity of codimension five (neglecting symmetry) \cite{golubitsky2012singularities}. 
 The full bifurcation diagram for the elastic beam with spring support is shown in Fig.~\ref{pic:ext}. In all, the normal forms calculated analytically using the method of Liapunov-Schmidt reduction technique agrees with our numerical results. 

When we set $\zeta = 0$, we find that $g_{rrr}^-$ vanishes at  
\be
\left(\frac{\lambda}{\pi}\right)^2=\frac{16}{3}n^2.
\ee
This implies a critical load of 
\be
p^-=\frac{P^-}{P_E} = \frac{4}{3}n^2,
\ee
which agrees with earlier work by Magnusson et al. \cite{magnusson2001behaviour}.

\subsection{Effects of elastic foundation}


In our discussions thus far, we have focused on the case $\zeta=3$ (weak springs) used in our numerical calculations. As we vary $\zeta$, we find that $g_{r\delta p}^{+} < 0$  and $g_{r\delta p}^{-} > 0$ regardless of the value of $\zeta$. Likewise, $g_{rrr}^+$ is always positive for any $\zeta$. This means that we always observe supercritical pitchfork bifurcation at the critical loads $p^+_{\text{cr}}{(n)}$.

\begin{figure}[htb!]
\centering
\includegraphics[width=\textwidth]{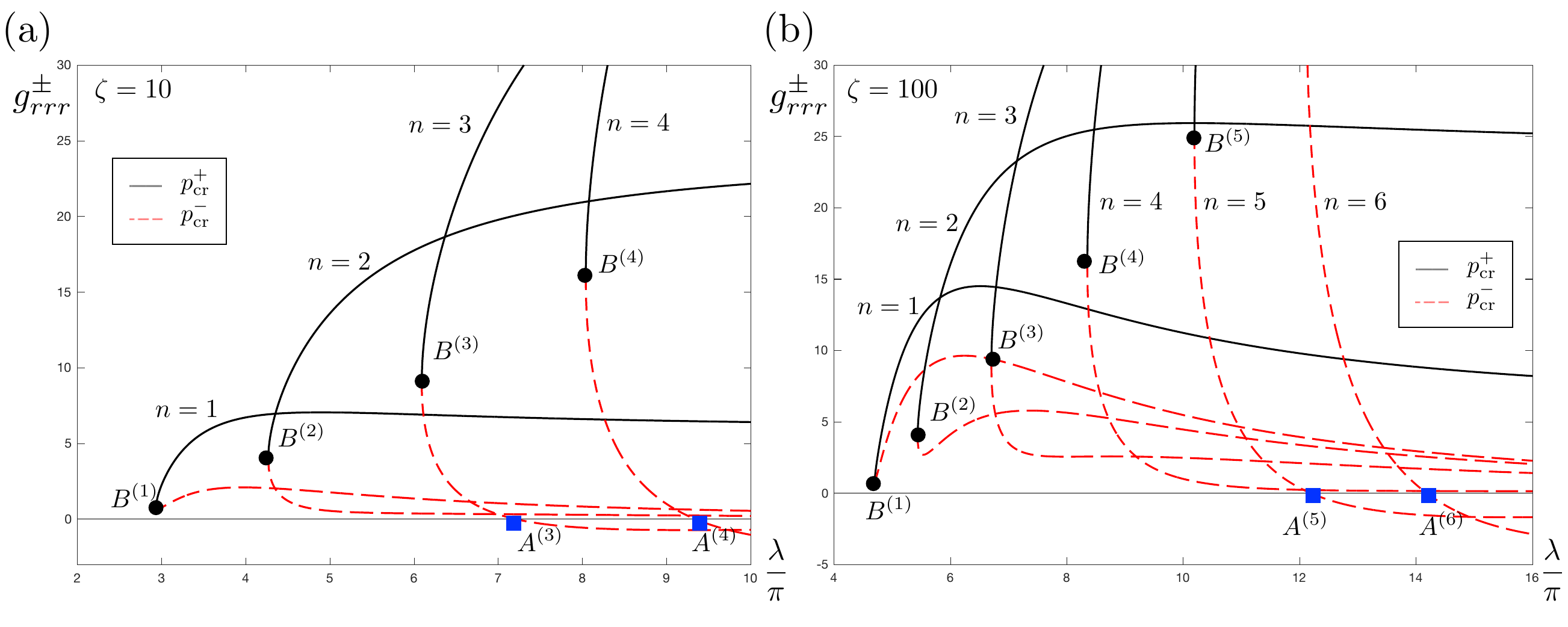}
\caption{Graph of $g_{rrr}^\pm$ (see Eq.~(\ref{eq:grrr})) as a function of slenderness $\lambda/\pi$. The solid lines denote $p^+$; the dashed lines denote $p^-$. (a) Elastic foundation with $\zeta=10$. (b) Elastic foundation with $\zeta=100$. }
\label{pic:Kappa}
\end{figure}

On the other hand, at the critical loads $p^-_{\text{cr}}{(n)}$, $g_{rrr}^-$ can change sign depending on the value of $\zeta$ and slenderness $\lambda$. At $\zeta=3$, $g_{rrr}^->0$ at $p^-_{\text{cr}}{(1)}$ for all slenderness $\lambda$ while for higher order modes ($n\ge2$), $g_{rrr}^->0$ if $ \lambda_c > \lambda \ge \lambda_0$ and $g_{rrr}^-\le 0$ if $\lambda \ge \lambda_c$. 
Consider the case $\lambda = 4.5\pi$: there are only two bifurcation modes ($n=1$ and 2) and $g_{rrr}^->0$ for both $n=1,2$ as evident from Figs.~\ref{pic:phase} and \ref{pic:recognition}. The beam undergoes subcritical pitchfork bifurcation at both $p^-_{\text{cr}}{(1)}$ and $p^-_{\text{cr}}{(2)}$. We find that the fundamental mode, i.e., straight beam, is stable until $p$ is above $p^+_{\text{cr}}{(2)}$ (rather than $p^-_{\text{cr}}{(1)}$ for the case without elastic support \cite{magnusson2001behaviour}). In this manner, we see that the elastic foundation
can help delay buckling and also promote buckling into $n=2$ mode, i.e., $\theta(x) \sim \cos(2\pi x/\ell)$, over the $n=1$ mode. 

At intermediate values of spring constants, say $\zeta=10$, we find that $g_{rrr}^->0$ at $p^-_{\text{cr}}$ for both $n=1$ and 2 for all slenderness $\lambda$ as shown in Fig.~\ref{pic:Kappa}(a). On the other hand, for the higher order modes, $g_{rrr}^-$ can change sign from positive to negative depending on whether $\lambda$ is smaller than or greater than $\lambda_c(n)$ (i.e., $A^{(n)}$). For example, at $\lambda = 7.5\pi$, there are three bifurcation modes ($n=1,2$ and 3) in total; $g_{rrr}^->0$ for $n=1$ and $2$, and $g_{rrr}^-<0$ for $n=3$. In this case, the beam undergoes subcritical pitchfork bifurcation at $p^-_{\text{cr}}{(1)}$ and $p^-_{\text{cr}}{(2)}$, and supercritical pitchfork bifurcation at $p^-_{\text{cr}}{(3)}$. This means that the elastic beam is stable until $p> p^-_{\text{cr}}(3)$ and will buckle preferentially into the $n=3$ mode, i.e., $\theta(x) \sim \cos(3\pi x/\ell)$, over the $n=1$ and 2 modes. 

At large values of spring constants, say $\zeta=100$, we find that $g_{rrr}^->0$ at $p^-_{\text{cr}}$ for $n=1,2,3,$ and 4 for all slenderness $\lambda$. For the higher order modes, $g_{rrr}^-$ can change sign from positive to negative depending on whether $\lambda$ is smaller than or greater than $\lambda_c(n)$. For example, at $\lambda = 10\pi$, there are a total of four bifurcation modes ($n=1,2,3$ and 4); $g_{rrr}^->0$ for $n=1, 2, 3$ and $4$ as illustrated in Fig.~\ref{pic:Kappa}(b). In this case, the beam undergoes subcritical pitchfork bifurcation at $p^-_{\text{cr}}{(1)}$, $p^-_{\text{cr}}{(2)}$, $p^-_{\text{cr}}{(3)}$, and $p^-_{\text{cr}}{(4)}$. This means that the elastic beam is stable until the axial load exceeds the critical value $p^+_{\text{cr}}(4)$ at which point it will buckle preferentially into the $n=4$ mode, i.e., $\theta(x) \sim \cos(4\pi x/\ell)$, instead of the $n=1,2$ and 3 modes. 

Suppose there are a total of $N$ permissible bifurcation modes at a given spring constant $\zeta$ and slenderness $\lambda$. Note that $N$ modes can lead to either $2N-1$ or $2N$ bifurcation points given by $p^\pm_{\text{cr}}(n)$ depending on whether there is a degenerate critical load ($p^+_{\text{cr}} = p^-_{\text{cr}}$) or not. 
The presence of the elastic foundation turn the first $M$ modes into subcritical pitchfork bifurcation. 
If $M<N$, then the first supercritical pitchfork bifurcation point is at $p^-_{\text{cr}}(M+1)$. 
This means that the elastic beam is stable until $p> p^-_{\text{cr}}(M+1)$ and will buckle preferentially into the $(M+1)$-th mode, i.e., $\theta(x) \sim \cos((M+1)\pi x/\ell)$. 
On the other hand, if $M=N$, then the first supercritical bifurcation point is at $p^+_{\text{cr}}(N)$. In this case, the straight beam is stable until $p>p^+_{\text{cr}}(N)$, at which point it buckles into the $N$-th mode, with a shape given by $\theta(x) = \cos (N\pi x/\ell)$. 
In both cases, we see that the presence of the elastic foundation causes the lowest few modes to turn into subcritical pitchfork bifurcations, leading to a higher effective critical load and a more undulated buckled shape. 
We can contrast this to an extensible beam without elastic foundation whereby the beam will buckle at either $p^+_{\text{cr}}(1)$ for $\lambda_0(1)<\lambda < \lambda_c(1)$ or $p^-_{\text{cr}}(1)$ for all $\lambda > \lambda_c(1)$. 
Thus, the presence of the elastic foundation makes it harder to buckle the elastic beam, increasing the critical load from $p^-_{\text{cr}}(1)$ to $p^-_{\text{cr}}(M+1)$ or $p^+_{\text{cr}}(N)$ depending on the values of $\zeta$ and $\lambda$. 
Overall, we find that the presence of the elastic foundation can lead to a myraid of buckled shapes for the classical elastica and one can tune the strength of the spring constant to generate the desired buckled modes---an experimentally testable prediction. 

%


\section{Conclusions}
\label{sec:conclusions}

In this work, we study the stability of an extensible elastic beam on a Winkler foundation under the action of a compressive force. $P$. 
Our model is based on the general Euler-Bernoulli theory of elastic beam and can be described by two 2nd order ordinary differential equations.
Earlier bifurcation work does not include the linear Hookean restoring force in the vertical direction \cite{magnusson2001behaviour}. 
The bifurcation analysis of the ODEs is calculated using the Liapunov-Schmidt reduction technique which leads to a single scalar equation involving a state variable and a bifurcation parameter \cite{golubitsky2012singularities}. 
The null space of the linearized differential equations has dimension one and on the basis of a single bifurcation equation, we obtain the bifurcation diagram which is very rich and there exist both supercritical and subcritical pitchfork bifurcations as well as other higher order bifurcations. 
There are two critical loads, $p^\pm_{\text{cr}}(n)$ for each bifurcation mode which depends on the slenderness of the beam and the spring constant in a nonlinear manner. All buckling modes at $p^+_{\text{cr}}(n)$ exhibit supercritical pitchfork bifurcation. 
The presence of the elastic foundation causes the lower order modes at $p^-_{\text{cr}}(n)$ to become subcritical pitchfork bifurcation and as a result, the first supercritical pitchfork bifurcation point occurs at a higher critical load. 
Thus the straight beam remains stable for a larger range of compressive loads due to the effects of the elastic foundation and when instability sets in, the beam buckles into a more undulated shape, i.e., higher order mode instead of $n=1$.
Our results are intuitive and conform with everyday experience. 
Our numerical calculations corroborate our analytical findings and we also observe post-buckling ``crossed'' fold bifurcation and transcritical bifurcation not reported before. 
Our framework imposes no restrictions on the size of the displacements and hence facilitates an full analysis of the entire postbuckling range.
In all, our study has uncovered the subtle interplay between elasticity and foundation effects in a minimal setting for an extensible beam with experimentally testable predictions.

\ack{

E.H.Y. acknowledged support from the Singapore Ministry of Education through the Academic Research Fund Tier 1 (RG140/22) and Academic Research Fund Tier 2 (MOE-T2EP50223-0014). L.M. acknowledges the support of the Simons Foundation and the Henri Seydoux Fund. 
 }

\section*{References}
\bibliographystyle{iopart-num-long}

\bibliography{references}

\providecommand{\newblock}{}
\begin{thebibliography}{10}
\expandafter\ifx\csname url\endcsname\relax
  \def\url#1{{\tt #1}}\fi
\expandafter\ifx\csname urlprefix\endcsname\relax\def\urlprefix{URL }\fi
\providecommand{\eprint}[2][arXiv]{#1:\linebreak[0]#2}

\bibitem{levien2008elastica}
Levien R 2008 The elastica: a mathematical history {\em Electrical Engineering
  and Computer Sciences University of California at Berkeley\/} {\bf 70}
  \urlprefix\url{https://www2.eecs.berkeley.edu/Pubs/TechRpts/2008/EECS-2008-103.pdf}

\bibitem{love2013treatise}
Love A~E~H 2013 {\em A treatise on the mathematical theory of elasticity\/} 4th
  ed (Cambridge university press)
  \urlprefix\url{https://www.cambridge.org/9781107618091}

\bibitem{antmanbook}
Antman S~S 2005 Problems in nonlinear elasticity {\em Nonlinear Problems of
  Elasticity\/}
  \urlprefix\url{https://link.springer.com/book/10.1007/0-387-27649-1}

\bibitem{hoyle2006pattern}
Hoyle R~B 2006 {\em Pattern formation: an introduction to methods\/} (Cambridge
  University Press)

\bibitem{todhunter2014history}
Todhunter I 2014 {\em A History of the Theory of Elasticity and of the Strength
  of Materials\/} vol~1 (Cambridge University Press)
  \urlprefix\url{https://www.cambridge.org/core/books/history-of-the-theory-of-elasticity-and-of-the-strength-of-materials/F1F62FAC50BB14FF166D16F337210744}

\bibitem{zaccaria2011structures}
Zaccaria D, Bigoni D, Noselli G and Misseroni D 2011 Structures buckling under
  tensile dead load {\em Proceedings of the Royal Society A: Mathematical,
  Physical and Engineering Sciences\/} {\bf 467} 1686--1700
  \urlprefix\url{https://royalsocietypublishing.org/doi/10.1098/rspa.2010.0505}

\bibitem{bigoni2012nonlinear}
Bigoni D 2012 {\em Nonlinear solid mechanics: bifurcation theory and material
  instability\/} (Cambridge University Press)
  \urlprefix\url{https://www.cambridge.org/core/books/nonlinear-solid-mechanics/16CBEA1E909A9CC2F11E70EFECBFB063}

\bibitem{van2024soft}
Van~Saarloos W, Vitelli V and Zeravcic Z 2024 {\em Soft Matter: Concepts,
  Phenomena, and Applications\/} (Princeton University Press)
  \urlprefix\url{https://press.princeton.edu/books/hardcover/9780691191300/soft-matter?srsltid=AfmBOoqUHB5-xEDm2dNPYPHBP9Lyz6aqFx5c6-G7SEo65NC7JMrq8D7s}

\bibitem{sahin2012physical}
Sahin O, Yong E~H, Driks A and Mahadevan L 2012 Physical basis for the adaptive
  flexibility of bacillus spore coats {\em Journal of The Royal Society
  Interface\/} {\bf 9} 3156--3160
  \urlprefix\url{https://royalsocietypublishing.org/doi/abs/10.1098/rsif.2012.0470}

\bibitem{magnusson2001behaviour}
Magnusson A, Ristinmaa M and Ljung C 2001 Behaviour of the extensible elastica
  solution {\em International Journal of Solids and Structures\/} {\bf 38}
  8441--8457
  \urlprefix\url{https://www.sciencedirect.com/science/article/pii/S0020768301000890?casa_token=rK4PIXGVPpgAAAAA:xosbNZj7eI9QeU_GZ_WSmoqWAWlUSWfYv8oW-ripabHld3fbKpbUbHmIzcW4frsCNqlELDcTJJ8}

\bibitem{golubitsky2012singularities}
Golubitsky M and Schaeffer D~G 1985 {\em Singularities and Groups in
  Bifurcation Theory: Volume I\/} (New York: Springer-Verlag)
  \urlprefix\url{https://link.springer.com/book/10.1007/978-1-4612-5034-0}

\bibitem{reissner1972one}
Reissner E 1972 On one-dimensional finite-strain beam theory: the plane problem
  {\em Zeitschrift f{\"u}r angewandte Mathematik und Physik ZAMP\/} {\bf 23}
  795--804 \urlprefix\url{https://link.springer.com/article/10.1007/bf01602645}

\bibitem{landau2012theory}
Landau L~D, Pitaevskii L, Kosevich A~M and Lifshitz E~M 2012 {\em Theory of
  elasticity: volume 7\/} 3rd ed (New York: Elsevier)
  \urlprefix\url{https://books.google.com/books?hl=en&lr=&id=NXRaWJb4HdkC&oi=fnd&pg=PP1&dq=Theory+of+elasticity:+volume+7+landau&ots=fT4uyZon1i&sig=5feX61Hmqp9sb9_jcLcNoozykjM}

\bibitem{keener2018principles}
Keener J~P 2018 {\em Principles of applied mathematics: transformation and
  approximation\/} (CRC Press)
  \urlprefix\url{https://www.taylorfrancis.com/books/mono/10.1201/9780429493263/principles-applied-mathematics-james-keener}

\bibitem{ciarlet1988three}
Ciarlet P~G 1988 {\em Mathematical Elasticity\/} ({\em Studies in Mathematics
  and its Applications\/} vol~1) (Amsterdam: Elsevier)

\bibitem{winkler1867lehre}
Winkler E 1867 Die lehre von elastizitat und festigkeit (the theory of
  elasticity and stiffness) {\em H. Domenicus. Prague\/}
  \urlprefix\url{https://archive.org/details/bub_gb_25E5AAAAcAAJ}

\bibitem{eisenberger1985exact}
Eisenberger M and Yankelevsky D~Z 1985 Exact stiffness matrix for beams on
  elastic foundation {\em Computers \& structures\/} {\bf 21} 1355--1359
  \urlprefix\url{https://www.sciencedirect.com/science/article/pii/0045794985901890}

\bibitem{thambiratnam1996dynamic}
Thambiratnam D and Zhuge Y 1996 Dynamic analysis of beams on an elastic
  foundation subjected to moving loads {\em Journal of sound and vibration\/}
  {\bf 198} 149--169
  \urlprefix\url{https://www.sciencedirect.com/science/article/pii/S0022460X96905623}

\bibitem{nicolau1982compressible}
Nicolau A and Huddleston J 1982 The compressible elastica on an elastic
  foundation {\em Journal of Applied Mechanics\/} {\bf 49} 577

\bibitem{michaels2019geometric}
Michaels T~C, Kusters R, Dear A~J, Storm C, Weaver J~C and Mahadevan L 2019
  Geometric localization in supported elastic struts {\em Proceedings of the
  Royal Society A\/} {\bf 475} 20190370

\bibitem{trefethen2017exploring}
Trefethen L~N, Birkisson {\'A} and Driscoll T~A 2017 {\em Exploring ODEs\/}
  (SIAM)
  \urlprefix\url{https://epubs.siam.org/doi/book/10.1137/1.9781611975161}

\bibitem{chen2020biocrust}
Chen N, Yu K, Jia R, Teng J and Zhao C 2020 Biocrust as one of multiple stable
  states in global drylands {\em Science Advances\/} {\bf 6} eaay3763
  \urlprefix\url{https://www.science.org/doi/abs/10.1126/sciadv.aay3763}

\bibitem{armitstead1996study}
Armitstead J, Bertram C and Jensen O~E 1996 A study of the bifurcation
  behaviour of a model of flow through a collapsible tube {\em Bulletin of
  mathematical biology\/} {\bf 58} 611--641
  \urlprefix\url{https://link.springer.com/article/10.1007/BF02459476}

\end{thebibliography}

\end{document}